\documentclass[conference]{IEEEtran}
\IEEEoverridecommandlockouts
\usepackage{cite}
\usepackage{amsmath,amssymb,amsfonts}
\usepackage{algorithmic}
\usepackage{graphicx}
\usepackage{textcomp}
\usepackage{xcolor}
\usepackage{subfigure}
\usepackage{multirow}
\usepackage{epstopdf}
\usepackage{stfloats}
\usepackage{multirow}

\def\BibTeX{{\rm B\kern-.05em{\sc i\kern-.025em b}\kern-.08em
    T\kern-.1667em\lower.7ex\hbox{E}\kern-.125emX}}
\DeclareRobustCommand*{\IEEEauthorrefmark}[1]{%
   \raisebox{0pt}[0pt][0pt]{\textsuperscript{\footnotesize\ensuremath{#1}}}}
\begin{document}
\title{Height-Dependent LoS Probability Model for A2G MmWave Communications under Built-up Scenarios}
\author{
\IEEEauthorblockN{Minghui~Pang\IEEEauthorrefmark{1}, Qiuming~Zhu\IEEEauthorrefmark{1,}\IEEEauthorrefmark{*}, Fei~Bai\IEEEauthorrefmark{1},
Zhuo~Li\IEEEauthorrefmark{2}, Hanpeng~Li\IEEEauthorrefmark{1}, Kai~Mao\IEEEauthorrefmark{1},Yue~Tian\IEEEauthorrefmark{1}}
\IEEEauthorblockA{\IEEEauthorrefmark{1}The Key Laboratory of Dynamic Cognitive System of Electromagnetic Spectrum Space, \\
College of Electronic and Information Engineering, \\
Nanjing University of Aeronautics and Astronautics, Nanjing 211106, China}
\IEEEauthorblockA{\IEEEauthorrefmark{2}The Key Laboratory of Radar Imaging and Microwave Photonics,Ministry of Education, \\
College of Electronic and Information Engineering, \\
Nanjing University of Aeronautics and Astronautics, Nanjing 211106, China}
\IEEEauthorblockA{\IEEEauthorrefmark{*}Corresponding author}
Email: \{pangminghui, zhuqiuming, baifei, lizhuo, sz2004013, maokai\}@nuaa.edu.cn\\
\{Pollon1999\}@163.com
}
\maketitle
\begin{abstract}
Based on the three-dimensional propagation characteristic under built-up scenarios, a height-dependent line-of-sight (LoS) probability model for air-to-ground (A2G) millimeter wave (mmWave) communications is proposed in this paper. With comprehensive considerations of scenario factors, i.e., building height distribution, building width, building space, and the heights of transceivers, this paper upgrades the prediction method of International Telecommunication Union-Radio (ITU-R) standard to both low altitude and high altitude cases. In order to speed up the LoS probability prediction, an approximate parametric model is also developed based on the theoretical expression. The simulation results based on ray-tracing (RT) method show that the proposed model has good consistency with existing models at the low altitude. However, it has better performance at the high altitude. The new model can be used for the A2G channel modeling and performance analysis such as cell coverage, outage probability, and bit error rate of A2G communication systems.
\end{abstract}
\section{Introduction}
\IEEEPARstart{U}{nmanned}  Aerial Vehicles (UAVs) have shown great growth in various fields due to their high mobility, low cost, and easy deployment\cite{Zhu18_chinacom}. MmWave communications have the advantages of fast transmission rate, high security, and spatial multiplexing of spectrum resources\cite{Cheng20_ITJ}. The UAV-aided air-to-ground (A2G) mmwave communication technology is promising for the beyond fifth and sixth generation (B5G/6G) mobile communication systems\cite{Li19_ITJ}. It is well-known that the A2G channel has exhibited different characteristics from traditional communication channels due to 3D scattering environments and valid scatterers only around ground\cite{Zhu18_trans, Khawaja19_CST}. Especially, the line-of-sight (LoS) path dominates the reliability of A2G communication link but it appears randomly due to UAV's high mobility and maneuver\cite{Zhu19_IET}. Therefore, it is vital to explore and build a LoS probability model for channel modeling and performance evaluation of A2G communication systems. Note that A2G channels in this paper are also suitable for the airship, air balloon, and other aircrafts to the grounds.

There are limited theoretical or measurement studies in the literatures for the LoS probability modeling\cite{Zhu21_3GPP, ITU-R2135, WINNER, 5GCM, Samimi15_WCL, Lee18_TAP, Jarvelainen16_WCL, ITU-R1410, Holis08_TAP, Hourani14_WCL, Hourani20_WCL, Liu18_CL, Cui20_ITJ}. Several standardized channel models, e.g., the Third Generation Partnership Project (3GPP) TR 38.901\cite{Zhu21_3GPP}, International Telecommunication Union-Radio (ITU-R) M.2135-1\cite{ITU-R2135}, WINNER II\cite{WINNER} and 5G Channel Model (5GCM)\cite{5GCM}, have their own LoS probability calculation methods, but most of them are based on the measurement results and only suitable for the low-altitude (or traditional mobile) scenarios. Moreover, these empirical models can provide accurate results for specific environments but difficult to be extended to other environments with different distributions of buildings. Since channel measurements are expensive and time consuming, some researchers studied the LoS probability based on the simulation data, obtained from the ray-tracing (RT) method\cite{Samimi15_WCL, Lee18_TAP} and point cloud \cite{Jarvelainen16_WCL}. However, a prior accurate knowledge of environment information is required for this kind of deterministic methods, which is unrealistic for most of applications.

The analytical method based on the geometry can provide acceptable estimation with simplicity and generality \cite{ITU-R1410, Holis08_TAP, Hourani14_WCL, Hourani20_WCL, Liu18_CL, Cui20_ITJ}. It describes the built-up scenarios by several statistical parameters and derived the LoS probability based on the geometric relationships. For example, a widely accepted analytical method was studied in the ITU-R Rec. P.1410 model \cite{ITU-R1410}, but the factor of building width was not included. By conducting simulations for massive locations of mobile terminal, the LoS probability function was fitted respect to the elevation angle in \cite{Holis08_TAP, Hourani14_WCL}. Note that these two models were only suitable for the high-altitude case in which the height was greater than 1000 m. The authors in \cite{Hourani20_WCL} modeled the buildings by random distributed cylinders with random heights. The height obeyed log-normal distribution and the distribution of buildings followed the Poisson point process by analyzing the data of Melbourne city. The authors in \cite{Liu18_CL, Cui20_ITJ} introduced the frequency factor into the LoS probability model by considering the frequency-related beam width. The method was general but it's complicated to identify the LoS component within the LoS clearance zone.

This paper aims to find a general LoS probability model for A2G mmWave communications with a good tradeoff between complexity and accuracy. The proposed model is suitable for both low and high altitudes, and includes the geometric factors such as the locations of terminals, building height distribution, building width, and building space. In addition, the properties of new LoS probability model are investigated and a simplified parametric model for deriving the closed-form solutions of channel parameters and system performance is also derived, which can greatly reduce the computational complexity with good accuracy.

The remainder of paper is organized as follows. Section II gives the definition of LoS probability under built-up scenarios. In Section III, the theoretical model is derived and a corresponding simplified parametric model is also obtained. The simulation results and validations are given in Section IV. Finally, conclusions are drawn in Section V.

\begin{figure*}[!b]
\setcounter{equation}{7}
\begin{equation}
{{P}_{\text{LoS}}}({{d}_{\text{RX}}},{{h}_{\text{TX}}},{{h}_{\text{RX}}},\psi )=\prod\limits_{i=\text{1}}^{{{N}_{b}}}{{{P}_{i}}}
= \prod\limits_{i=\text{1}}^{{{N}_{b}}}{\left[ 1-\exp \left( -\dfrac{{{\left[ {{h}_{\text{TX}}}-\left( \dfrac{i-0.5}{{{N}_{b}}}\text{+}\dfrac{W}{\text{2}{{d}_{\text{RX}}}} \right)({{h}_{\text{TX}}}-{{h}_{\text{RX}}}) \right]}^{2}}}{2{{\gamma }^{\text{2}}}} \right) \right]}
\label{8}
\end{equation}
\vspace{-2ex}
\end{figure*}
\section{A2G Propagation Scenarios}
The built-up scenario is one of the most common application scenarios for the future A2G communications, where buildings are referred as main obstacles. In order to achieve a more general LoS probability prediction method, the environment-dependent geometrical parameters are usually described in a stochastic way. The main characteristic of built-up area is the layout of buildings. A typical A2G communication link under built-up scenario is shown in Fig. 1. In the figure, ${{h}_{\text{TX}}},{{h}_{\text{RX}}}$ represent the heights of UAV as the transmitter (TX) and vehicle as the receiver (RX), respectively and ${{d}_{\text{RX}}}$ is the horizontal distance from the TX to the RX.
\begin{figure}[!b]
	\centering
	\includegraphics[width=80mm]{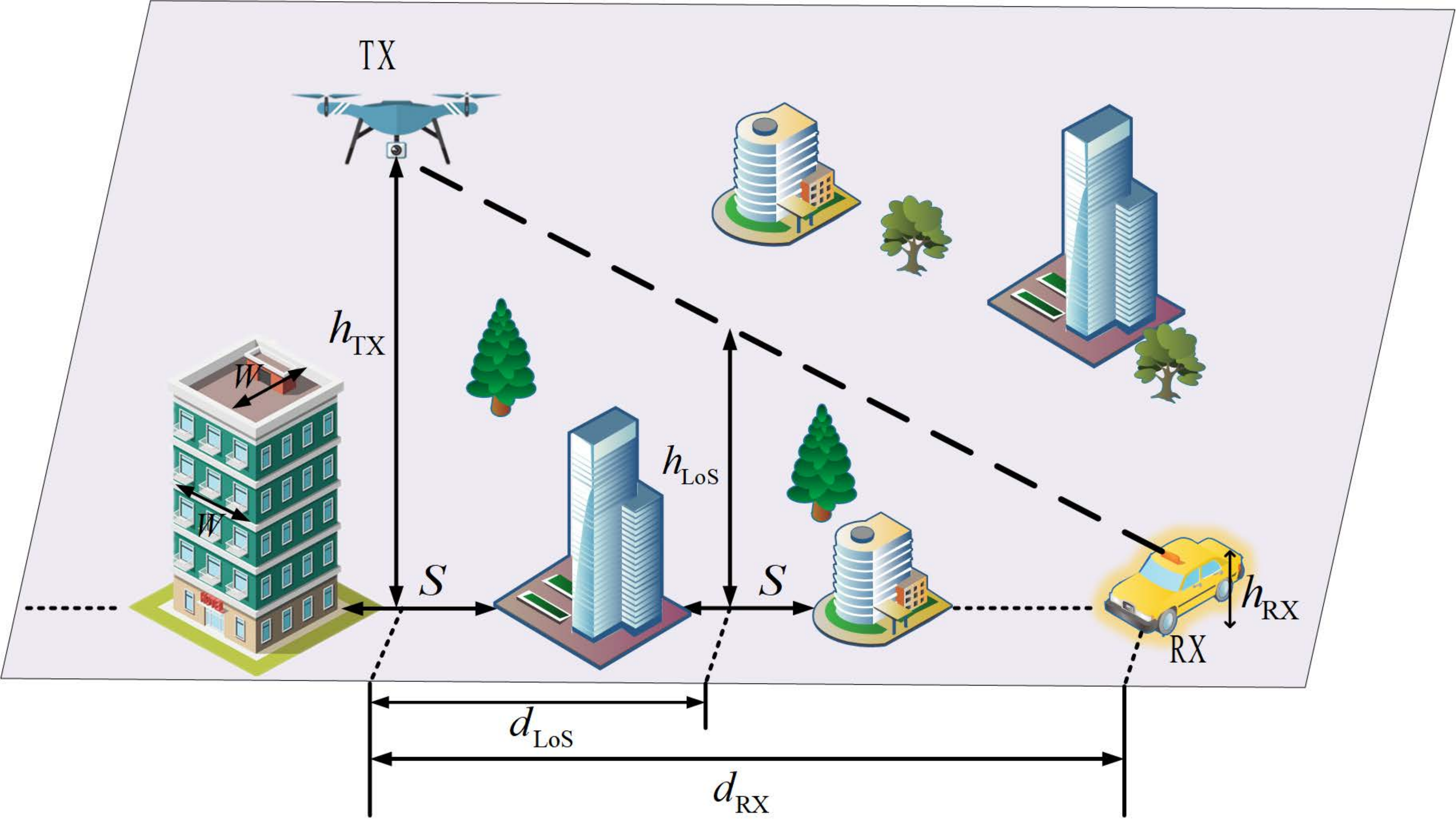}
	\caption{LoS propagation under built-up scenarios.}
    \label{fig:1}
\end{figure}

A well-known classification for built-up scenarios, i.e., suburban, urban, dense urban and high-rise urban can be addressed in\cite{Holis08_TAP}. It describes different scenarios by three parameters $\psi \in \{\alpha ,\beta ,\gamma \}$ defined in\cite{ITU-R1410}, where $\alpha $ is the percent of land area covered by buildings, $\beta $ represents the mean number of buildings per unit area, and $\gamma $ is a random variable denoting the random building height with the probability density function (PDF) as
\setcounter{equation}{0}
\begin{equation}
{P(h)=\frac{h}{{{\gamma }^{\text{2}}}}\exp \left( -\frac{{{h}^{2}}}{2{{\gamma }^{2}}} \right)}.
\label{1}
\end{equation}
\noindent In Fig. 1, $S,W$ represent the building space and width which can be obtained from the real data or the above scenario-dependent parameters as
\begin{equation}
{W = 1000\sqrt {\alpha /\beta }}
\label{2}
\end{equation}
\begin{equation}
{S = (1000/\sqrt \beta  )(1 - \sqrt \alpha  )}.
\label{3}
\end{equation}
\noindent The environment-dependent parameters of four typical scenarios are shown in the Table I.
\begin{table}[!t]
\renewcommand{\arraystretch}{1.4}
\caption{Environment parameters for different built-up scenarios.}
\label{table 1}
\centering
\begin{tabular}{p{1.8cm}p{0.6cm}p{0.6cm}p{0.6cm}p{0.6cm}p{0.6cm}}
\hline
Scenarios & $\alpha $ & $\beta $ & $\gamma $ & $W$ & $S$  \\
\hline
Suburban & 0.1 & 750 & 8 & 11.55 & 24.97\\
Urban& 0.3 & 500 & 15 & 24.49 & 20.23\\
Dense~urban& 0.5 & 300 & 20 & 40.82 & 16.91\\
High-rise~urban& 0.5 & 300 & 20 & 40.82 & 16.91\\
\hline
\end{tabular}
\end{table}
\section{Height-Dependent LoS Probability Model}
\subsection{Theoretical LoS Probability Derivation}
It should be mentioned that during the propagation of radio waves, radio radiation is not only concentrated on the optical path. The Fresnel zone is the area where the electric field strength is affected by obstacles. Theoretically, only when the Fresnel zone is completely blocked by obstacles can the LoS propagation be cut off. As shown in Fig. 2, we find that at a certain distance, the width of Fresnel zone decreases as the frequency increases and increases as the distance increases. In addition, the width of Fresnel zone at the mmWave band is within 2.5 m, but the distance between the receiver and transmitter  is several hundred meters. Therefore, for the mmWave frequency band, the width of Fresnel zone can be approximately ignored.
\begin{figure}[!b]
	\centering
	\includegraphics[width=80mm]{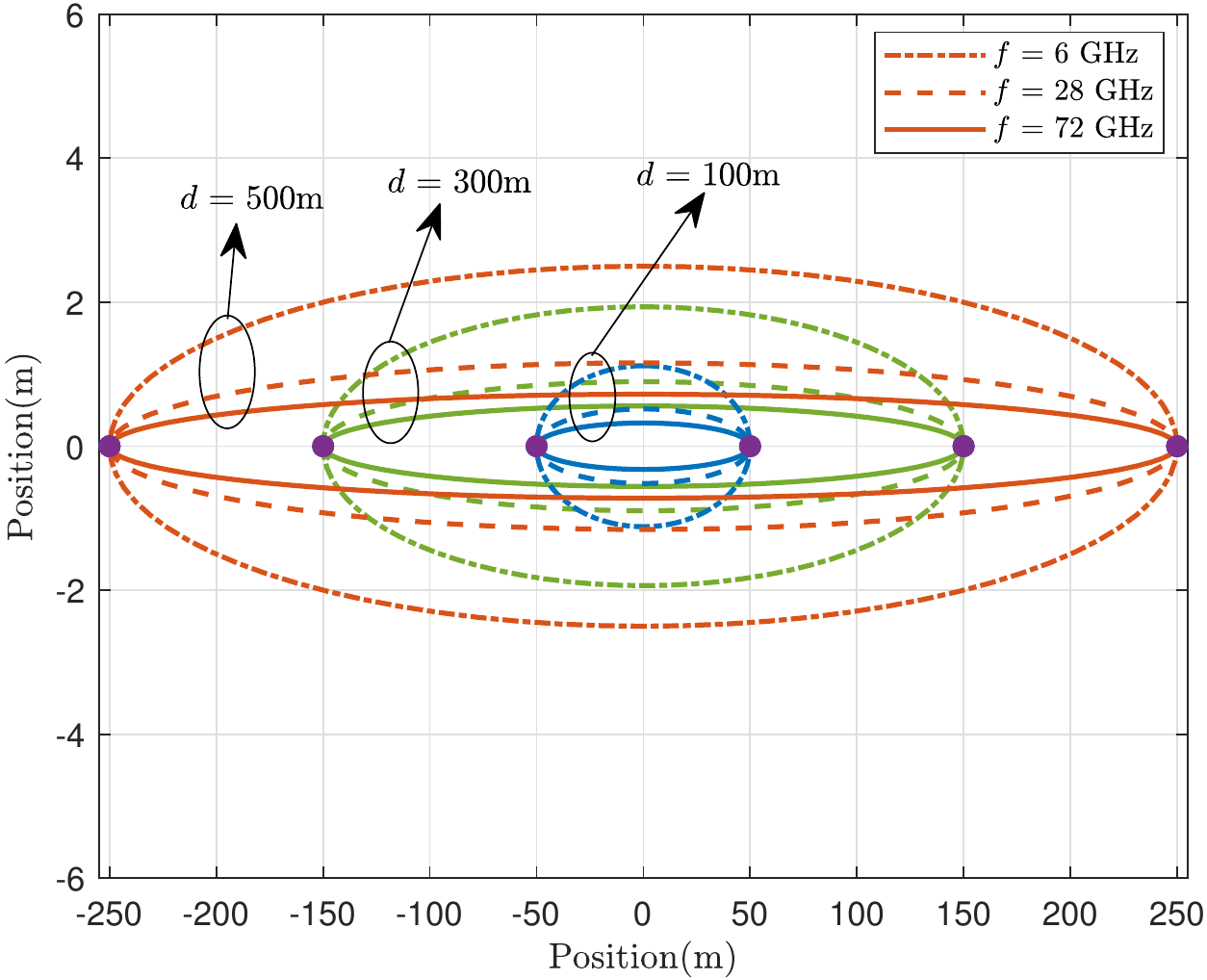}
	\caption{Fresnel zone profile with different distances and frequencies.}
    \label{fig:2}
\end{figure}
\begin{figure}[!t]
	\centering
	\includegraphics[width=80mm]{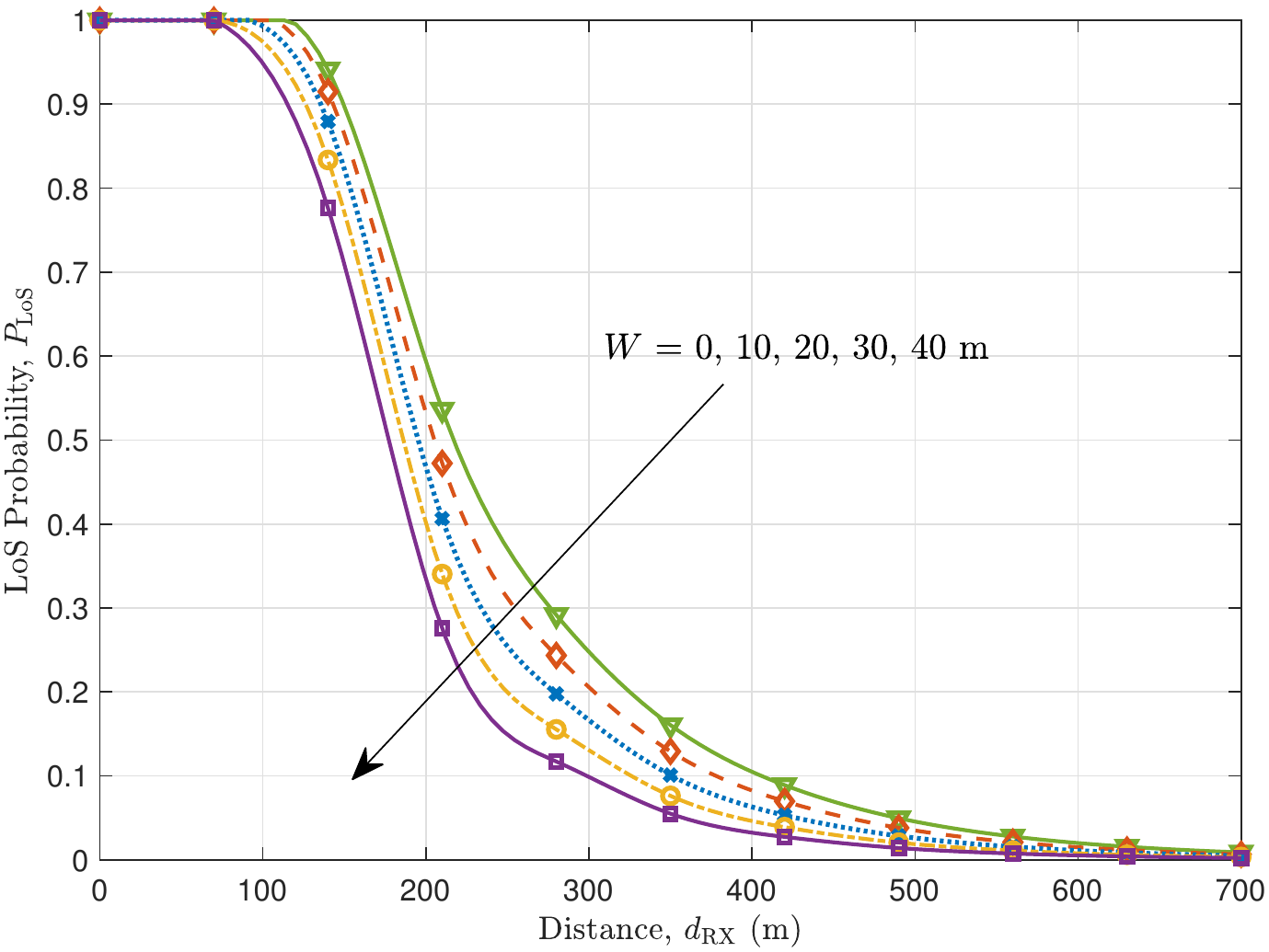}
	\caption{LoS probability with different building width.}
    \label{fig:3}
\end{figure}
When the positions of TX and RX are known, the probability of LoS path can be viewed as the probability that all building along the propagation path are below the height of connection line between the TX to the RX. The height of connection line at any point can be proved as
\begin{equation}
{{h_{{\rm{LoS}}}} = {h_{{\rm{TX}}}} - \frac{{{d_{{\rm{LoS}}}}({h_{{\rm{TX}}}} - {h_{{\rm{RX}}}})}}{{{d_{{\rm{RX}}}}}}}
\label{4}
\end{equation}
where ${{d}_{\text{LoS}}}$ is the distance from the TX to the obstruction. Furthermore, the LoS probability can be defined as
\begin{equation}
{{P_{{\rm{LoS}}}} = \prod\limits_{i = {\rm{0}}}^{{N_b}} {{P_i} = } \prod\limits_{i = {\rm{0}}}^{{N_b}} {P({h_i} < {h_{_{{\rm{LoS}}}}})}}
\label{5}
\end{equation}
\noindent where ${{h}_{i}}$ is the building height. Assuming that all buildings and streets are equally distributed, the number of buildings ${{N}_{b}}$ can be calculated as
\begin{equation}
{{N_b} = {\rm{floor}}\left( {\frac{{{d_{{\rm{RX}}}}\sqrt \alpha  }}{{W + S}}} \right){\rm{ = floor}}\left( {\frac{{{d_{{\rm{RX}}}}\sqrt {\alpha \beta } }}{{{\rm{1000}}}}} \right)}
\label{6}
\end{equation}
\noindent where the function floor represents the meaning of rounding down. The visibility from the TX to the RX means that the LoS path is not blocked by any building between the transceivers. Based on the environmental parameters, the probability ${{P}_{i}}$ that a building height is smaller than height ${{h}_{\text{LoS}}}$ can be obtained by
\begin{small}
\begin{equation}
{{P_i} = P({h_i} < {h_{_{{\rm{LoS}}}}}) = \int_0^{{h_{_{{\rm{LoS}}}}}} {P(h){\rm{d}}h = 1 - \exp \left( { - \frac{{h_{{\rm{LoS}}}^2}}{{2{\gamma ^2}}}} \right)}}.
\label{7}
\end{equation}
\end{small}
\noindent The LoS probability is obtained weighting each ${{P}_{i}}$ with weights $i$ dependent on the distance from the transmitter. It accounts for the number of buildings being greater at larger distance. By substituting (7) to (5), we can get the theoretical result related to width and height which can be expressed as (8).
\noindent As it shows that the LoS probability is fully determined by the position and environmental parameters. It should be mentioned that the new model is independent on the frequency and applicable for any heights of TX and RX. Moreover, the factor of building width is taken into account and the special case of  $W=0$ reduces to the same result as eq. (4) in \cite{Hourani14_WCL}.

In order to observe the effect of building width on the LoS probability, Fig. 3 shows the LoS probability with the central frequency $f$ = 28 GHz, ${{h}_{\text{TX}}}=70\text{ m}$, ${{h}_{\text{RX}}}=1.5\text{ m}$ and $\gamma =15$ at four different building width. It clearly shows that the LoS probability declines when building width increases. The reason is that the wider buildings would increase the block possibility.

\subsection{Simplified Parametric Expression}
The analysis model proposed in this paper can be applied to A2G scenarios with different building heights, and also takes the environment information such as the width of the houses into account. However, due to the continuous multiplication process, the calculation of the LoS probability is slightly complicated. For solving this problem and deriving a closed-form solution of system performance and channel model, we need a simple and closed-form expression of LoS probability. In order to obtain a closed-form expression and reduce the computational complexity, we further approximate the theoretical model by a parametric simplified model for specific scenarios.

Noted that the LoS probability is normally modeled as exponential functions against the distance, which ${{D}_{1}}$ is the breakpoint distance where LoS probability is no longer equal to 1, and ${{D}_{2}}$ is a decay parameter. Based on the proposed theoretical model, we describe ${{D}_{1}}$ and ${{D}_{2}}$ as height-dependent parameters for universality and applicability. The approximate model can be obtained by the minimum mean square error (MMSE) criteria as (9),
\begin{figure*}[!t]
\setcounter{equation}{8}
\begin{equation}
\begin{array}{l}
{{P}_{\text{LoS}}}({{d}_{\text{RX}}},{h}')=\min \left( \dfrac{{{D}_{1}}}{{{d}_{\text{RX}}}},1 \right)\cdot \left[ 1-\exp \left( -\dfrac{{{d}_{\text{RX}}}}{{{D}_{2}}} \right) \right]+\exp \left( -\dfrac{{{d}_{\text{RX}}}}{{{D}_{2}}} \right)\\
\ \ \ \ \ \ \ \ \ \ \ \ \ \ \ \ \ =\min \left( \dfrac{{{a}_{1}}{{{{h}'}}^{{{b}_{1}}}}+{{c}_{1}}}{{{d}_{\text{RX}}}},1 \right)\cdot \left[ 1-\exp \left( -\dfrac{{{d}_{\text{RX}}}}{{{a}_{2}}{{{{h}'}}^{{{b}_{2}}}}} \right) \right]+\exp \left( -\dfrac{{{d}_{\text{RX}}}}{{{a}_{2}}{{{{h}'}}^{{{b}_{2}}}}} \right) \\
\end{array}
\label{9}
\end{equation}
\end{figure*}

\noindent where ${h}'={{h}_{\text{TX}}}-{{h}_{\text{RX}}}$ and ${{d}_{\text{RX}}}$ is the horizontal distance from the TX to the RX. Note that the LoS probability is described by the distance and height difference where the ground plane is viewed as the height of ${{h}_{\text{RX}}}$. Then, we link the model parameters to the height-dependent variables. The optimal parameters for four typical built-up scenarios are summarized in Table II.
\begin{table}[!t]
\renewcommand{\arraystretch}{1.4}
\caption{Model parameters for different built-up scenarios.}
\label{table 2}
\centering
\begin{tabular}{p{1.8cm}p{0.8cm}p{0.8cm}p{0.8cm}p{0.8cm}p{0.8cm}}
\hline
Scenarios & $a_1$ & $b_1$ & $c_1$ & $a_2$ & $b_2$\\
\hline
Suburban & 1.698 & 1.082 & 30.07 & 38.63 & 0.4911\\
Urban & 0.3891 & 1.098 & 23.92 & 21.31 & 0.4770\\
Dense~urban & 0.3475 & 1.018 & 20.15 & 18.87 & 0.4461\\
High-rise~urban & 0.1885 & 0.9723 & 17.31 & 15.70 & 0.4106\\
\hline
\end{tabular}
\end{table}
\section{RT-based Validation and Comparison}
\subsection{RT-based Simulation Method}
The RT technique has been widely adopted to radio channels modeling, alleviating the burden of measurement campaigns. In the RT simulation, the electromagnetic wave departing from the source is considered to be a bunch of rays using a ray-optic approximation, so a geometric solution can be obtained based on the uniform theory of diffraction and geometric optics. In addition, the ray-optic approximation method well describes the short-wavelength electromagnetic wave propagation, so the RT technique has good performance for the small specific area and high frequency. After tracking all rays with the forward or the reverse technique, a lot of propagation parameters can be obtained. In this paper, we only focus on the LoS probability characteristic.

The flowchart of LoS probability prediction with RT method is shown in Fig. 4. The proposed algorithm in this paper mainly includes two steps, i.e., reconstruction of scenario and probability computation by RT techniques. In this paper, we obtain the width, height and space of the buildings according to the environmental parameters $\psi $ and reconstruct four built-up scenarios. In each scenario, we set up $N$ transmitters of different heights. Afterwards taking TX as a center, the positions of RX are distributed in concentric circles where the distance between the transmitter and receiver is equal.

In this paper, the simulations are performed at the central frequency of 28~GHz, with a bandwidth of 500 MHz. In order to obtain LoS probability data at different communication heights, we set TX heights starts from 0 m to 1000 m with the step of 10 m and the RX height is fixed at 2 m. The rest simulation parameters are described in Table III.
\begin{figure}[!t]
	\centering
	\includegraphics[width=80mm]{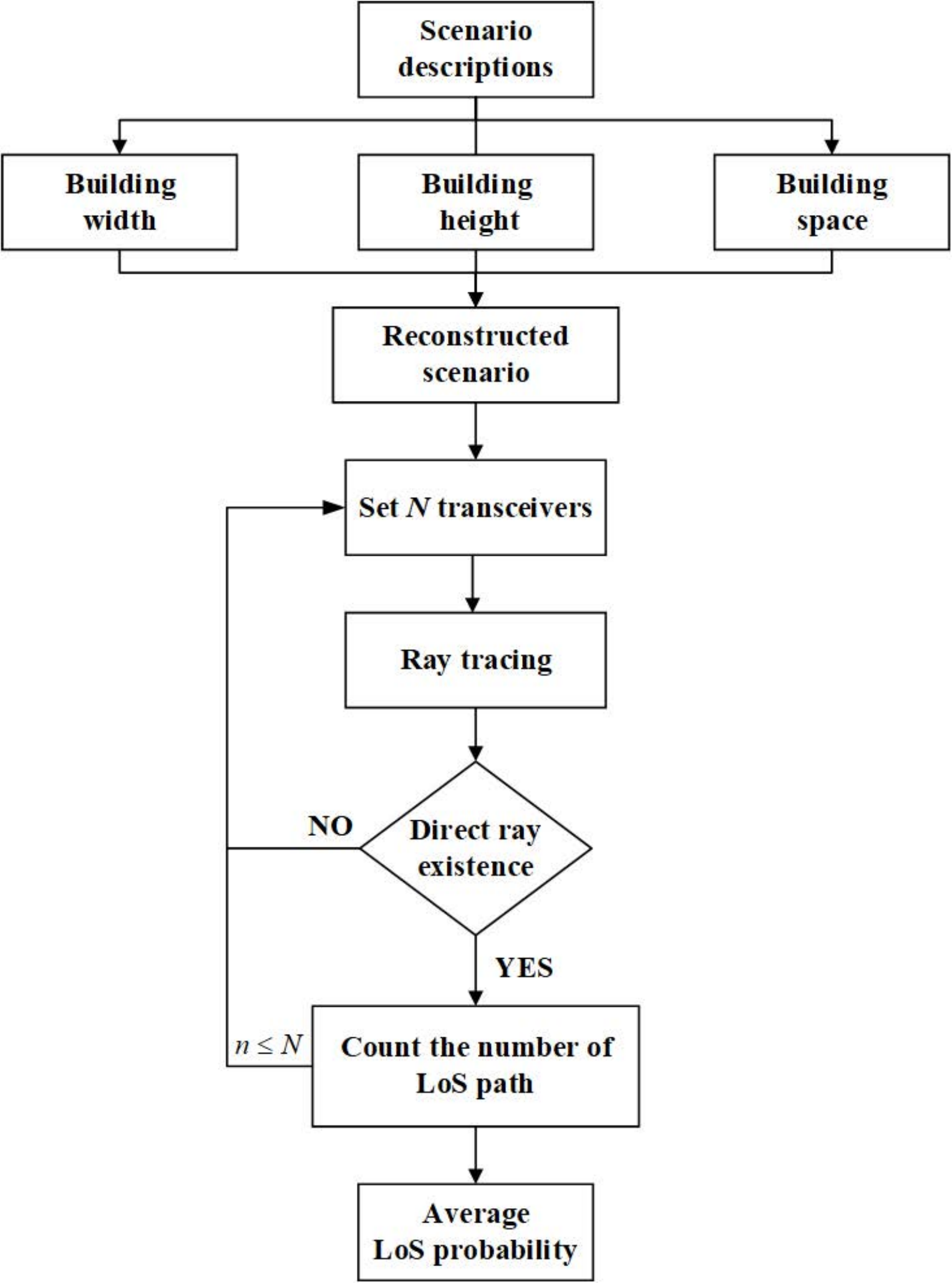}
	\caption{Flowchart of LoS probability simulation method.}
    \label{fig:4}
\end{figure}
\subsection{Validation and Comparison}
\begin{table}[!t]
\renewcommand{\arraystretch}{1.4}
\caption{Parameters of simulations.}
\label{table 3}
\centering
\begin{tabular}{p{3.5cm}p{3cm}}
\hline
Parameter & Value \\
\hline
Frequency & 28 GHz\\
Bandwidth & 500 MHz\\
Antenna type & omnidirectional\\
TX height & 0 -- 1000 m\\
RX height & 2 m\\
RX distance & 0 -- 1000 m\\
\hline
\end{tabular}
\end{table}
\begin{figure}[!b]
	\centering
	\includegraphics[width=80mm]{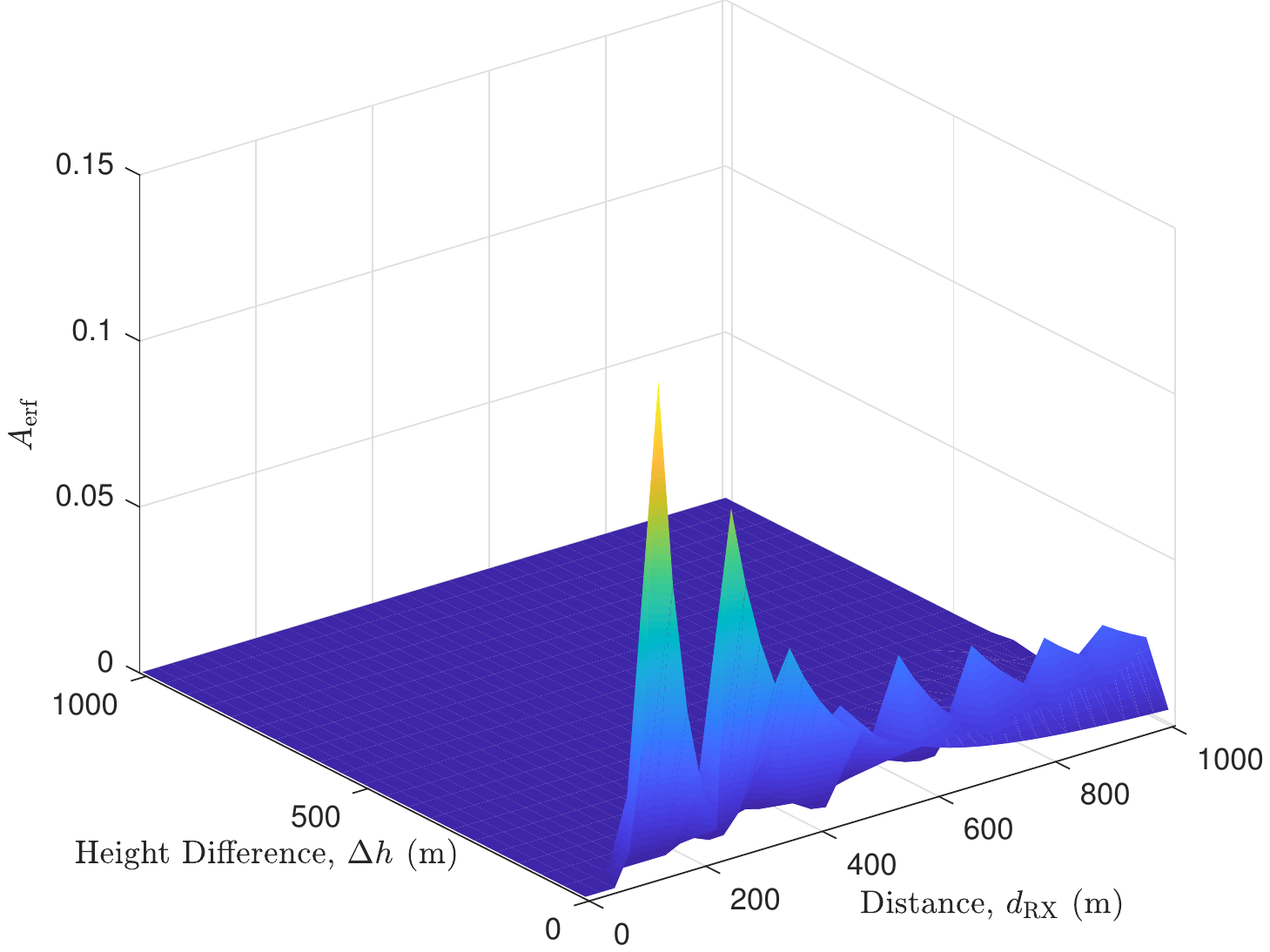}
	\caption{Absolute error function under suburban scenarios.}
    \label{fig:5}
\end{figure}
\begin{figure*}[!t]
\centering
\subfigure[Low altitude (30m)]{
\includegraphics[width=80mm]{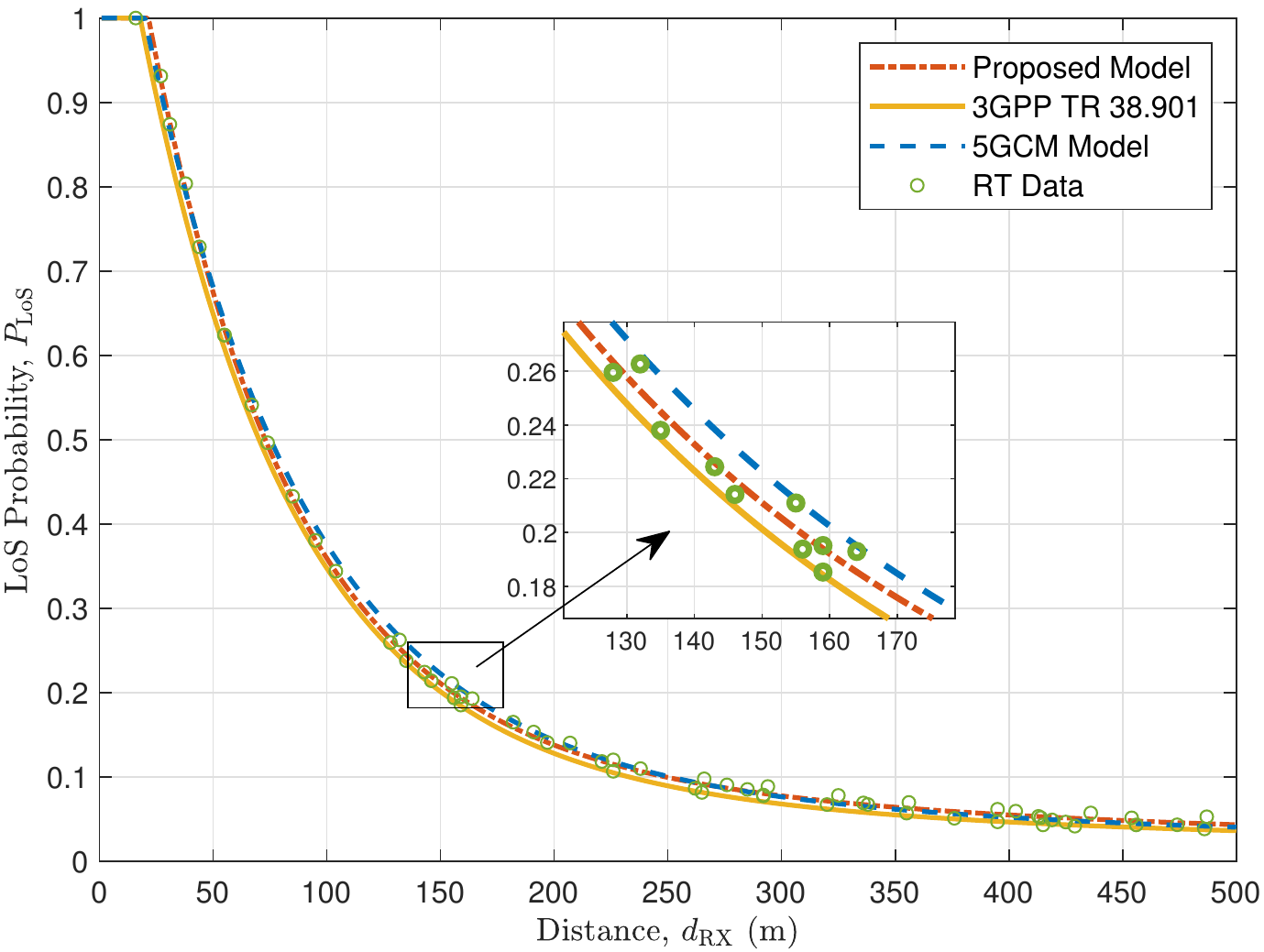}
}
\subfigure[High altitude (1000m)]{
\includegraphics[width=80mm]{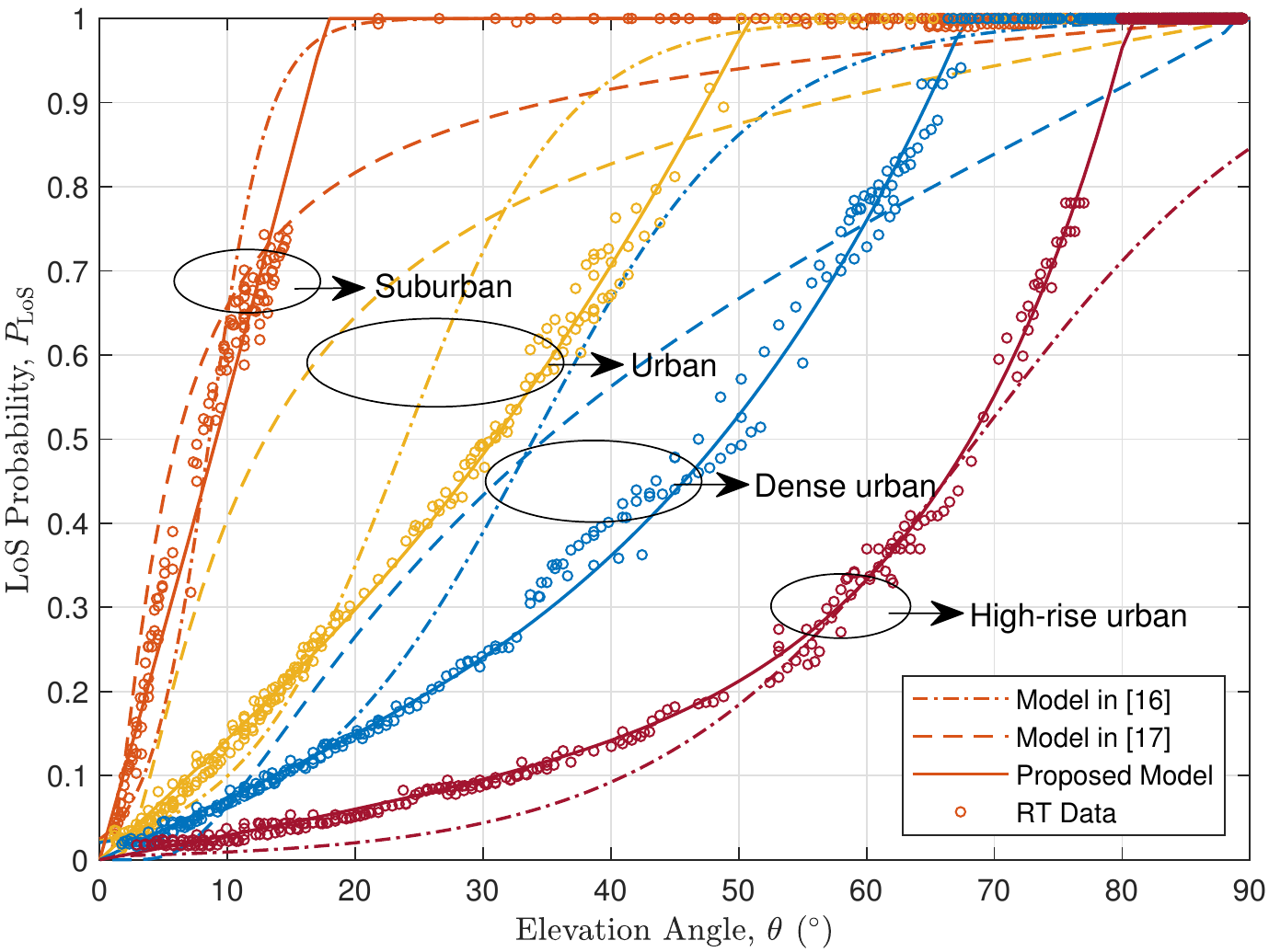}
}
\caption{Comparisons of different models with different heights.}
\end{figure*}
In this section, we run simulations for different heights and distances to theoretically validate the performance of proposed model of proposed model. For comparison purpose, we set ${{h}_{\text{RX}}}=0$ in the theoretical model to ensure that ${{h}_{\text{TX}}}$ and ${h}'$ represent the same physical meaning. In order to visually display the results, we set the absolute error function between the theoretical model and approximate model. It can be expressed as
\begin{equation}
{{A_{erf}} = {[{P_{{\rm{LoS}}}}({d_{{\rm{RX}}}},{h_{{\rm{TX}}}},{h_{{\rm{RX}}}},\psi ){\rm{  -  }}{P_{{\rm{LoS}}}}({d_{{\rm{RX}}}},h')]^2}}
\label{10}
\end{equation}
\noindent where ${{P}_{\text{LoS}}}({{d}_{\text{RX}}},{{h}_{\text{TX}}},{{h}_{\text{RX}}},\psi )$ is the probability of the theoretical model and the ${{P}_{\text{LoS}}}({{d}_{\text{RX}}},{h}')$ is the the probability of the approximate model. The absolute error simulation results are shown in Fig. 5 and the range of ${{d}_{\text{RX}}}$ and $\Delta h\text{=}{{h}_{\text{TX}}}-{{h}_{\text{RX}}}$ are from 0 m -- 1000 m. Fig. 5 shows that the absolute error between the analytical model and approximate model is from 0 -- 0.1. The good agreement between analytical and approximate results shows the applicability of simplified parametric model. In fact, this approximate model with lower complexity and closed-form expression may have wider application in the future channel modeling and communication performance evaluation.

In order to demonstrate the validity of proposed model in A2G scenarios, we also compare them with other previous models. Since measurement campaigns for A2G channels are difficult and high cost, we conduct tremendous RT simulations and obtained huge number of LoS probability data for comparison. Since some standard models such as 3GPP and 5GCM model are only suitable for low altitude, taking 30 m as an example, we compare our proposed model with them and RT data in Fig. 6(a). Good consistency across all models and RT data proves that our model can also achieve good prediction at low altitudes. For the high altitude case, taking 1000 m as an example, we compare our proposed model with other applicable models\cite{Hourani14_WCL, Hourani20_WCL}. For comparison purpose, we use the elevation angle $\theta =\arctan ({h}'/{{d}_{\text{RX}}})$ to describe the LoS probability as it did in\cite{Hourani14_WCL, Hourani20_WCL}. Fig. 6(b) shows that three models have similar trends but our proposed model is much closer to the RT data.
\section{Conclusions}
In this paper, we have derived a height-dependent LoS probability model based on the statistical characteristic and geometric parameters of built-up scenarios. In addition, a simplified parametric model has been proposed to speed up the prediction process as well as assist the closed-form solution derivation for channel model and system performance. The simulation results have shown that the proposed model demonstrates good versatility at both low and high altitudes and has a good agreement with RT-simulated data. The proposed model has a wide range of applications such as channel modeling, performance evaluation, and system optimization for A2G communications. Future work will include the channel measurement of LoS probability and incorporating the factor of antenna pattern.
\section*{Acknowledgment}
This work was supported by the National Key Scientific Instrument and Equipment Development Project (No.~61827801), Aeronautical Science Foundation of China (No.~201901052001), State Key Laboratory of Integrated Services Networks Funding (ISN22-11), and Open Foundation for Graduate Innovation of NUAA (No.~KFJJ20200416).


\vspace{12pt}
\end{document}